# Computational modeling of protein interactions and phosphoform kinetics in the KaiABC cyanobacterial circadian clock

Mark Byrne[1]

[1] Physics Department, Spring Hill College, 4000 Dauphin St., Mobile AL 36608

Corresponding author:

    Dr. Mark Byrne
    Physics Dept.
    4000 Dauphin St
    Spring Hill College
    Mobile, AL  36608  USA
    TEL: 251-380-3080
    Email: mbyrne@shc.edu

Abstract
The KaiABC circadian clock from cyanobacteria is the only known three-protein oscillatory system which can be reconstituted outside the cell and which displays sustained periodic dynamics in various molecular state variables. Despite many recent experimental and theoretical studies there are several open questions regarding the central mechanism(s) responsible for creating this ~24 hour clock in terms of molecular assembly/disassembly of the proteins and site-dependent phosphorylation and dephosphorylation of KaiC monomers. Simulations of protein-protein interactions and phosphorylation reactions constrained by analytical fits to partial reaction experimental data support the central mechanism of oscillation as KaiB-induced KaiA sequestration in KaiABC complexes associated with the extent of Ser$^{431}$ phosphorylation in KaiC hexamers  A simple two-state deterministic model in terms of the degree of phosphorylation of Ser$^{431}$ and Thr$^{432}$ sites alone can reproduce the previously observed circadian oscillation in the four population monomer phosphoforms in terms of waveform, amplitude and phase. This suggests that a cyclic phosphorylation scheme (involving cooperativity between adjacent Ser$^{431}$ and Thr$^{432}$ sites) is not necessary for creating oscillations. Direct simulations of the clock predict the minimum number of serine-only monomer subunits associated with KaiA sequestration and release, highlight the role of monomer exchange in rapid synchronization, and predict the average number of KaiA dimers sequestered per KaiC hexamer.

Introduction
The cyanobacterial circadian clock is perhaps the simplest autonomous circadian oscillatory system that can be reconstituted in vitro, requiring just the three proteins KaiA, KaiB and KaiC and ATP (1). The KaiABC post-translational oscillator (PTO) has been hypothesized to be the central oscillator in vivo in cyanobacteria since the PTO functions in constant darkness while protein and mRNA are not rhythmic (2), exogenous inhibition of KaiC protein and mRNA synthesis does not alter the circadian periodic dynamics of the PTO (2), and various mutations in key phosphorylation sites in KaiC result in a loss of sustained circadian rythmicity as measured by bioluminescence of cellular populations (3,4). In vivo, a clock model consisting of a dual oscillatory mechanism including a transcription-translation feedback loop (TTFL) in addition to a PTO has been proposed (5); it has also been suggested that a dual-oscillator system is necessary under rapid growth conditions (6). An alternative model suggests that the PTO is the central oscillator for circadian rhythmicity and the TTFL is a driven oscillator which dampens absent a sustained KaiC phosphorylation rhythm (7).

In vitro the oscillations in phosphorylation levels are sustained without damping for more than 10 days in non-limiting ATP (8). The central core protein, monomeric KaiC, self-assembles into a homo-hexamer (9) and KaiC both auto-phosphosphorylates and auto-dephosphorylates (10,11). Based on macroscopic population studies, dimeric KaiA enhances the rate of phosphorylation and inhibits the dephosphorylation rate of KaiC (12). Tetrameric KaiB abrogates the effect of KaiA, typically resulting in population dephosphorylation of KaiC (11). Mutational studies identified two primary phosphorylation sites on the KaiC monomer, Serine-431 (S$^{431}$) and Threonine-T432 (T$^{432}$) (4,14). The population monomeric phosphoform abundances oscillate in a circadian manner with T$^{432}$ phosphorylation phase leading S$^{431}$ phosphorylation (15,16). The population phosphoform data, including mutational studies, has been interpreted in the context of a 4-state model approximated by the population monomer transitions, unphosphorylated → threonine-only phosphorylated → doubly phosphorylated → serine-only phosphorylated: (S,T) → (S,pT) → (pS,pT) → (pS,T) (16). Based on mutational studies this sequence was further hypothesized to operate at the individual monomer level (15) rather than simply illustrating population kinetics, suggesting that the rates of phosphorylation and dephosphorylation of S$^{431}$ and T$^{432}$ were sensitive to the presence or absence of phosphates on the adjacent amino acid.

Various mathematical models of the clock have taken the approach of modeling the population phosphoform distributions without considering the statistical distribution of hexamer states and without explicitly including protein-protein interactions (e.g, ref. 16) . Other models use as state variables the concentration of KaiC hexamers with different degrees of phosphorylation in the form $[C_i](t)$, with $i$ labeling degree of phosphorylation, and model protein-protein interactions using mass-action kinetics. In these models the kinetics of phosphorylation and dephosphorylation are represented using transition rates between hexamer states (7,17-23). Within the former models one cannot ask questions related to hexameric states, stoichiometry, nor predict the dynamics of various protein complexes over the cycle, which is likely essential for understanding the in vivo clock. In the latter mathematical models the number of potential distinct hexamer configurations is large, in principle consisting of 84 unique states resulting from 6 monomers with 4 different phosphorylated forms possible for each monomer (23). Furthermore, each unique form of the KaiC hexamer may differentially interact with KaiA and KaiB. Additional simplifications of models are therefore generally used to reduce the relevant number of KaiC hexamer states to simulate and probe mechanisms of oscillation in the system (see for example, ref. 18 or the comprehensive hybrid model of ref. 23). No mathematical model to date has simulated the kinetics of KaiC hexamers interacting with KaiA and KaiB, explicitly including the complete stochastic statistical distribution of hexamers and taking into account both $S^{431}$ and $T^{432}$ phosphorylation and dephosphorylation reactions. Because the KaiC hexamer, rather than the individual monomers, interacts with KaiA and KaiB, understanding the system in terms of hexamer interactions and including the distribution of hexamer states is central to investigating the mechanism of synchrony.

To address a variety of questions related to protein-protein interactions and the generation of a sustained oscillatory dynamics a previous stochastic matrix model of the system (24) was updated to include both $S^{431}$ and $T^{432}$ sites on each monomer for individual hexamers (e.g., a matrix of $N$ hexamers by 12 columns), with phosphorylation and dephosphorylation at these sites simulated using Monte-Carlo in which 0s and 1s represent the absence or presence of phosphates on the $S^{431}$ and $T^{432}$ sites. The numerical matrix model has the advantage that it automatically includes the complete statistical distribution of KaiC hexamer states, immediately probes stochasticity, can be used to test alternative scenarios for protein binding or release from KaiC, and can be used to unambiguously study the effect of monomer exchange among hexamers on clock dynamics.

In the first section the partial reaction kinetics for the phosphorylation phase (KaiA+KaiC) and dephosphorylation phase (KaiC alone, KaiB +KaiC) are used as a constraint and it is shown that the phosphoform experimental data can be fit analytically using simple effective first order site-independent phosphorylation and dephosphorylation rates on $S^{431}$ and $T^{432}$. Inserting these population rates in the matrix model, except now operative at the single monomer/subunit level, reproduces the analytical curves up to stochastic effects. The matrix model with the constrained rates is used to predict the predicted quasi-asymptotic ( > 20 hrs) distribution of phosphates across hexamers for the phosphorylation phase (KaiA + KaiC). While the partial reaction kinetics are useful for characterizing the phosphorylation and dephosphorylation phases separately, they provide little to no information on the mechanism(s) required for switching from one phase to another with respect to protein-protein interactions and population synchronization. In the second section various potential scenarios in matrix simulations are examined that produce, or do not produce, a functional clock. KaiA sequestration is assumed to be the primary mechanism for phase synchronization in the model as suggested by previous experimental data and theoretical models (16-18,23). We analyze varying assumptions for KaiB binding and KaiA sequestration, and varying assumptions for the release of KaiA and KaiB from KaiABC complexes. The model predicts quantitative circadian protein abundance oscillations for the complexes and these predicted abundances are compared to previous experimental data; in addition phase mixing of samples for the standard concentrations are simulated and compared to data. Finally varying protein concentrations in the model and the effect on circadian dynamics are discussed.

**Results**
**The monomer phosphoform kinetics from partial reaction data (KaiA+ KaiC, KaiC alone) are consistent with a simple two-state model.**
A simple two-state model of site-dependent phosphorylation and dephosphorylation is sufficient to qualitatively and quantitatively fit the partial reaction data for the population monomer phosphoforms as previously reported in Refs. 15 and 16. We use the notation (pS,T), (S,pT) and (pS,pT) to indicate the population monomer phosphoform abundances and S and T to indicate the abundance of phosphorylated $S^{431}$ and $T^{432}$ sites, respectively. The phosphorylation phase (KaiA +KaiC) and dephosphorylation phase (KaiC alone or KaiC + KaiB) are modeled using simple first order kinetics. The assumption in each phase is that the reactions proceed as simple first order reactions with constant effective rates. In the phosphorylation phase, fixed KaiA and KaiC concentrations are assumed (the standard in vitro concentrations) which results in constant effective phosphorylation and dephosphorylation rates in the population for each site; in general these rates depend on the chosen initial KaiA and KaiC concentrations. It can be shown that the time-dependence of the fractional occupancy of sites from 1$^{st}$ order kinetics with phosphorylation rate $k_+$ and dephosphorylation rate $k_-$ starting from an initially un-phosphorylated state is

(1)     *f(t) = [1/(1 + $k_-/k_+$)] {1 - exp [- ( $k_+$ + $k_-$) t] }*

We tentatively assume this function is valid for describing the population of S and T during the phosphorylation phase (with different rates for $S^{431}$ vs $T^{432}$). Similarly, for the dephosphorylation phase, starting from max occupancy $f_0$, 1$^{st}$ order kinetics implies the following functional form for the fractional occupancy:

(2)     *f(t) = (1/(1 + $k_-/k_+$)) { 1 - [1 - (1 + $k_-/k_+$) $f_0$] exp [- ($k_+$ + $k_-$) t] }*

These relations, Eqns. *(1)* and *(2),* describe the kinetics of the degree of phosphorylation of the individual sites ($S^{431}$ and $T^{432}$) in the population, but not the monomer kinetics. From simple probabilistic arguments applied to a well-randomized population and assuming site-independency, the fraction of serine-only monomers is $f_{(pS,T)} = f_S*(1- f_T)$, the fraction of doubly-phosphoryled monomers (pS,pT) is $f_{(pS,pT)} = f_S*f_T$ and that of threonine-only (S,pT) monomers is $f_{(S,pT)} = f_T*(1- f_S)$. Therefore even though the population phosphoform kinetics for the individual sites is very simple (Fig 1B), when monomer abundances are measured in terms of (pS,T), (S,pT) and (pS,pT) the kinetics for the monomers appear to be more complex (the derivation and explicit analytical expressions for the monomer phosphoforms are given in text SI).

The analytical equations yields curves in good agreement with the experimental data from Refs. (15) and (16) for the KaiA + KaiC partial reaction data (Fig 1A). What appears to be a transition from threonine-only to doubly-phosphorylated monomers in the population is a probabilistic effect related to a slower $S^{431}$ phosphorylation rate compared to the $T^{432}$. Analogous analytical solutions for the dephosphorylation phase (KaiC alone) also produce good fits to previous experimental phosphoform data (Fig 1A), where a faster rate for $T^{432}$ dephosphorylation yields a transient increase in serine-only monomers. This suggests, in terms of phosphorylation rates, that the period of the oscillator is primarily sensitive to the effective rate of $S^{431}$ phosphorylation (since this is the "slow" variable) during the phosphorylation phase and to the effective $S^{431}$ dephosphorylation rate during the dephosphorylation phase.

To study the partial reactions at the hexamer level a matrix model was constructed with a fixed number of hexamers and explicitly including the 12 sites (6$S^{431}$ and 6$T^{432}$ sites) for each hexamer. Phosphorylation and dephosphorylation of sites is implemented probabilistically (transitions of 0 → 1 or 1 → 0) where the macroscopic rates are now interpreted in the context of a fixed probability of reaction per computational timestep. Using the rates determined from fits with the analytical equations, the matrix simulations for the phosphoforms reproduce the same curves for both the phosphorylation and dephosphorylation phases (up to stochastic effects) as expected (Fig 1A); implementing exchange of monomers among hexamers does not affect the curves except for minor smoothing. The matrix model can be used to specify the corresponding phosphorylation distribution of hexamers as a function of time (KaiA+KaiC); after adding KaiA to KaiC at t ~ 20 hrs the model predicts that the majority of hexamers have 7-9 phosphates bound with very few hexamers having all 12 bound (Fig 1C). The average KaiC hexamer in the population at the end of the phosphorylation phase has a monomer subunit distribution consisting of 3(pS,pT), 2(S,pT) and either 1(S,T) or 1(pS,T) monomer.

**Description of the 2-state model with free-KaiA mediated rates**

The partial reaction data give insufficient information about the system to generate oscillations because they do not characterize the mechanism(s) used in the two population transitions, the phosphorylation to dephosphorylation phase transition and vice versa. In order for the KaiC population to enter a dephosphorylating phase (under the standard in vitro conditions) KaiB is required. Experimental data suggest KaiB preferentially binds to hexamers which are hyper-phosphorylated and are either abundant in (pS,pT) and/or (pS,T) monomers (22). The specific monomeric distributions in hexamers that specify KaiB's relative association and dissociation rates from KaiC are, however, not known experimentally. KaiB appears to render KaiA catalytically inactive, and we assume this occurs in tandem with the sequestration of KaiA in KaiB-KaiC complexes as previously suggested (16); no free "inactive" KaiA molecules have been experimentally observed, the proportion of KaiA in complexes appears to to stay at a high, almost constant level throughout the cycle (25), and the exogenous addition of KaiA during the middle of the dephosphorylation phase results in a rapid rise to population hyperphosphorylation (16). As KaiA is sequestered the effective phosphorylation and dephosphorylation rates at each site vary due to a decreased binding probability of KaiA to KaiC. We take this into account by allowing time-varying phosphorylation and dephosphorylation rates for the $S^{431}$ and $T^{432}$ sites ($f_S$ labels the fraction of $S^{431}$ sites and $f_T$ the fraction of $T^{432}$ sites phosphorylated in the population at time t)

(3)
$$df_S/dt = \tilde{k}_{+S}(1-f_S) - \tilde{k}_{-S} f_S$$
$$df_T/dt = \tilde{k}_{+T}(1-f_T) - \tilde{k}_{-T} f_T$$

The phosphorylation and dephosphorylation rates are assumed to depend hyperbolically on KaiA concentration (16):

(4)
$$\tilde{k}_i = k_i^0 + k_i^A f_A/(f_A + K')$$

where $i$ refers to $S^{431}$ or $T^{432}$ sites ($i = +/- S$ or $+/- T$), $f_A$ is the fraction of "free" non-sequestered KaiA dimers at time t relative to initial hexameric KaiC ( $[KaiC_0]$), $k_i^0$ is the rate when $f_A = 0$ (auto-phosphorylation and auto-dephosphorylation) and $k_i^A$ parameterizes the effect of KaiA on phosphorylation and dephosphorylation of the $S^{431}$ and $T^{432}$ sites. Experimentally it has been suggested that $K' = K_{1/2}/[KaiC_0] \sim 1/3$ using a standard concentration of approximately 0.6μM hexamer KaiC and

$K_{1/2} \sim 0.2\mu M$ as a function of KaiA dimer concentration (from supplement of Ref (16), which plotted the inferred $T^{432}$ phosphorylation rate versus monomeric [KaiA]). The eight rate parameters $k_i^0$ and $k_i^A$ are already fixed by fitting the analytic solutions to the partial reaction data when $f_A = 0$ and $f_A = 1$ (see text SI). The primary assumption in Eqn. (4) is that the rates of phosphorylation and dephosphorylation of individual sites are regulated by KaiA with the same functional hyperbolic dependence and the same $K_{1/2}$. Since the phosphorylation and dephosphorylation rates are altered by KaiA concentration (15,16, Fig 1) the functional form of Eqn (4) follows from assuming the rates depend on the probability that KaiA-KaiC bound states are formed.

These fixed rates for $S^{431}$ and $T^{432}$ phosphorylation and dephosphorylation were implemented in the matrix model and used to examine varying assumptions for KaiB binding to individual hexamers. We assume $N_s$ KaiA dimers on average are instantaneously sequestered following KaiB association with a KaiC hexamer. This assumption of instantaneous sequestration is supported by experiments indicating rapid (on a circadian timescale) KaiB-initiated dephosphorylation (16). Previous experimental data also suggested between *3* or *4* KaiA dimers are sequestered by a hexamer (25). For most matrix simulations an initial test value of *4* was assumed and was then varied to test for the presence or absence of oscillatory behavior. Simulations were used to study different scenarios for KaiB binding to a KaiC hexamer such that KaiB bound to a single hexamer when there was (a) a minimum specified number (*m*, 0 < *m* < 6) of serine-only, threonine-only, or doubly phosphorylated monomer subunits on the hexamer or (b) or a minimum total number of phosphates independent of monomer composition. The same scenarios were tested for KaiA release from an individual hexamer with the exception that the threshold specified release when the phosphoform or total phosphate number fell below a minimum threshold. There are only two rates in simulations which are not directly constrained by the experimental data, the effective KaiB binding rate (which includes the rate associated with sequestering $N_s$ KaiA dimers) and the rate of KaiA release. To exclude model tuning these were both set to occur at a similar rate to the constrained (maximal) phosphorylation and dephosphorylation reaction rates in the model (0.5 molecule$^{-1}$ hr$^{-1}$ for binding and 0.5 hr$^{-1}$ for KaiA release). There are no additional freely adjustable parameters in the model.

**Selective KaiB binding, KaiA sequestration, and selective KaiA release from KaiC hexamers generates experimentally realistic sustained circadian oscillations**
Circadian oscillatory dynamics do not occur for any hexameric phosphorylation threshold when only net phosphorylation is used as a criteria for KaiB binding-KaiA sequestration, and subsequent KaiA and KaiB release from KaiABC complexes (Fig 2A). This supports the necessity of a clock protein with at least two unique phosphorylatable residues; the scenario tested includes KaiA sequestration with mediation of rates (Eqns 3-4) and differential KaiB binding and release, and yet is incapable of generating a functional clock (using the constrained rate parameters). The physical interpretation of numerical simulations is as follows: $T^{432}$ sites are rapidly phosphorylated and $S^{431}$ sites more slowly (Fig 1B). KaiB does not bind until the $S^{431}$ population is sufficiently high and this can be implemented by binding to hexamers with sufficient serine-only monomers or serine phosphates or by binding to hexamers whose total phosphate number is sufficiently high (e.g., > 7, containing multiple (pS,pT) subunits). Binding and release based on total phosphate number alone does not produce sustained oscillations due to hexamer population heterogeneity at both phase transitions resulting in desynchronization. High amplitude circadian oscillatory dynamics occur when serine-only monomeric subunits on the KaiC hexamer are responsible for both binding and release of KaiA (Fig 2B). Simulations suggest *3* serine-only monomers on a KaiC hexamer as a minimum threshold for KaiB binding (Fig 2B) while higher thresholds for KaiB binding are not preferred due to their low abundance at the end of the phosphorylation phase (Fig 1C). The resulting phosphoform dynamics (Fig 2C) are a good match to the previous experiments of Refs (15) and (16). Oscillations also do not occur if total phosphate number alone is used for the release of KaiA and KaiB from the hexamer, even if three serine-only monomers are assumed as a threshold for KaiB binding. This indicates desynchronization of the population occurs in the dephosphorylation to phosphorylation phase transition if there is sufficient hexamer heterogeneity during either phase transition; in contrast, a serine-monomer based release threshold and the separation of timescales for dephosphorylation of the $S^{431}$ and $T^{432}$ sites assures that most $T^{432}$ sites have been dephosphorylated across the population near the end of the dephosphorylation phase assuring a relatively homogenous hexamer population at the transition.

The system can oscillate without monomer exchange (with a longer period) and also shows robust oscillations if one serine-only subunit is used for KaiA and KaiB release from the KaiABC complex. Several experiments indicate monomers are exchanged among hexamers throughout the cycle (8,24, 25, 19). We assume the models including exchange are preferred and use exchange in subsequent sections, in particular for phase mixing simulations (see below). With monomer exchange the binding of KaiB to hexamers with multiple serine-only monomer subunits results in an autocatalytic effect where serine-only monomers "shuffle" among hexamers initiating further KaiB binding (and KaiA sequestration) which catalyzes further creation of serine-only monomers due to varying population phosphorylation and dephosphorylation rates. The effect of monomer exchange is most obvious in the early dephosphorylation phase where the abundance of serine-only monomers rapidly rises; dominant (rapid) exchange during the early dephosphorylation phase has been reported previously in experiments (8) and supports the proposed autocatalytic mechanism. Monomer exchange is not required to generate oscillations from the structure of the model, however, as KaiA sequestration alone autocatalytically results in higher phosphorylated $S^{432}$ (and lower $T^{431}$) resulting in further KaiA sequestration. In addition to the role of serine-only monomers, the existence of oscillations clearly depends on the average number of sequestered KaiA dimers per KaiC hexamer ($N_s$) which, along with the $K' = K_{1/2}/[KaiC_0]$, determines the extent of nonlinearity in KaiA sequestration and release.

Motivated by the matrix model a very simple approximate 2-state deterministic model with sequestration of serine-only monomers can generate surprisingly realistic circadian phosphoform oscillations (Fig 2D). The model is given by Eqns. (3) and (4) with an "effective" *instantaneous* sequestration term for the fraction of free KaiA dimers via $N_s$ serine-only monomers:

(5) $\quad f_A(t) = max\{0, f_{Ai} - N_s f_{(pS,T)}\} = max\{0, f_{Ai} – N_s [fs (1-f_T)]\}$

where $f_{Ai}$ is the initial fraction of KaiA dimer to KaiC hexamer: $f_{Ai} =[KaiA]_0/[KaiC]_0$ (~1 for standard concentrations). The inversion formula between sites and monomers was used for a well-mixed (randomized) population. Realistic oscillations are found for $N_s \sim 4.2$, the predicted average number of serine-only monomers that sequester KaiA. The deterministic model is a rather drastic approximation as there is no explicit KaiB binding term, no KaiA release term, and the heterogeneous hexamer states are not included; $K'$ is therefore treated as an effective parameter in the deterministic model; $K' \sim 0.1$ or less is required for oscillations as was previously found by a 4-state deterministic model with serine-monomer sequestration (16). It is nonetheless surprising that the time-dependent partial reaction dynamics (KaiA + KaiC, KaiC alone) and phosphoform kinetics (KaiA + KaiB + KaiC) are well described by this simple 2-state deterministic model with one "adjustable" parameter based on free KaiA mediated rates of the $S^{431}$ and $T^{432}$ sites and sequestration associated with serine-only monomers.

**The proposed model is in good agreement several experimental constraints for standard concentrations**
In addition to accurately simulating the phosphoform population circadian oscillations the model can quantitatively predict the KaiBC/KaiABC protein complex kinetics and free KaiC levels in terms of abundance, phase, and waveform throughout the circadian cycle for the standard concentrations (Fig 3A). Allowing KaiB release from KaiC simultaneous with KaiA release and using the serine-monomer thresholds as previously described creates KaiBC/KaiABC dynamics which are approximately in phase with the serine-only monomer kinetics, as was observed experimentally (16); the relative phase and amplitude of KaiABC complexes can be modified by allowing KaiB release at a different rate than KaiA or by choosing a different release threshold. The predicted abundance variation in complexes is also in reasonable agreement with previous experiments (24,25). Due to multiple KaiA dimer sequestration per hexamer, most of the KaiC hexamers are in non-complexed form throughout the cycle with approximately 1/4 of KaiC hexamers having at least one KaiA bound. Ito et al (8) examined the mixing of multiple hexamer populations with variable phase, and experiments indicated the resulting KaiC population was dominated by the sample that was in the dephosphorylating phase. Simulations from the matrix model illustrate similar results to experiment (Fig 3B and 3C); the predominance of the dephosphorylation phase is interpreted in the model as due to the previously described exchange of serine-only monomers which drives autocatalytic KaiB binding and KaiA sequestration in the population. As the exchange of monomers is phase-independent all the hexamers are rapidly infused with serine-only monomers from the dephosphorylating sample.

Previous experiments have also investigated the effect of varying the initial relative concentrations of KaiA, KaiB or KaiC and the overall scaling of concentration relative to the standard values (25,26). The monomer concentrations for the standard reaction (1) are (1.2A, 3.5B, 3.5C) µM or (0.6$A_2$, 0.875$B_4$, 0.58$C_6$) µM or expressed relative to initial KaiC hexamer: (1.03 $A_2$, 1.51 $B_4$, 1$C_6$). The model as formulated above incompletely describes KaiA association and sequestration due to varying binding probabilities of KaiA and KaiB to KaiC since the effect of KaiA is solely encoded (for the standard concentrations) in the phosphorylation and dephosphorylation rates of the two sites (Eqn 4) and sequestration is implemented instantaneously with KaiB-KaiC complex formation. Nonetheless the model can reproduce complete loss of oscillations for both lower (0.25 x) and higher (3x) initial KaiA dimer/KaiC hexamer ratios resulting in hyperphosphorylation for high initial KaiA and hypophosphorylation for low values (Fig 4A). The interpretation is clear from the model; for low initial KaiA the $S^{431}$ and $T^{432}$ phosphorylation rates are too low to reach the threshold for KaiB association. For too high an initial KaiA concentration insufficient KaiA is sequestered and the sites remain hyperphosphorylated.

The model can reproduce the affect of both increasing and reducing the initial KaiB concentration (Fig 4B), keeping KaiA and KaiC fixed, provided a hyperbolic dependence of the rate of KaiB association with KaiC hexamers as a function of KaiB concentration is assumed (see SI text). The initial KaiB concentration enters the model in two places, in the association rate of KaiB with hexamers containing the requisite threshold number of serine monomers and when KaiB is limiting so the system is unable to sequester sufficient free KaiA in KaiABC complexes. Not including the effect of altered KaiB binding rates, the mechanism of sequestration alone suggests limiting KaiB is relevant when $f_B < f_A /3.5$ (both scaled to intial KaiC) and implies limiting KaiB when $f_B < 0.29$ indicating that oscillations should cease around ~1/3 [KaiB]$_0$ of the original standard concentration. For high initial KaiB concentrations there is little affect on oscillations (25-26) suggesting that for the standard concentration the KaiB assocation rate is near the maximal rate and higher concentrations only slightly increase the B-association rate with very little change in the system's dynamics consistent with simulations (Fig 4B).

The effect of overall concentration scaling on the dynamics is due to changes in the binding probability of KaiA and KaiB to KaiC. In the model above the effect of an overall concentration change in KaiA, KaiB and KaiC simultaneously is parameterized by $K' = K_{1/2}/[KaiC_0]$ in Eqn (4) and in the hyperbolic dependence of KaiB association with KaiC. The relationship between the free KaiA fraction and rates were fitted using the standard concentrations; the term $f_A/(f_A+0.3)$ can be interpreted as altering the maximal rates, $k_i^A$, in proportion to the fraction of the hexamer population with KaiA bound. The maximum rates ($k_i^A$) should also scale with simultaneous concentration changes in KaiA and KaiC (with scale factor γ, γ ≡ 1 for standard concentrations). A simple assumption is a hyperbolic dependence of these maximal rates on the overall scaling factor $k_i^{Anew}=k_i^A \gamma/(\gamma+\beta)$ with β << γ with the same affect of fractional KaiA occupancy on the maximal rate. Simulations (Fig 4C) verify that elevated concentration scaling of all proteins produce almost no changes in the kinetics (β =0.05 shown) while significantly reducing all protein concentrations results in longer periods and loss of oscillations. The interpretation from the model is that for the standard concentrations the rates are near their maximum values and are not altered much by further protein increases; furthermore sequestration via KaiB is unaltered as the ratio of KaiA to KaiB is unchanged. For significantly reduced concentrations of all proteins the primary effect is a decrease in the effective maximal rates for the S and T sites which results in loss of oscillations. These results are at least in qualitative agreement with Kageyama et al (25).

**Discussion**

We have presented the first simple deterministic 2-state model of the KaiABC oscillator displaying realistic phosphoform circadian oscillations in which the state variables of the oscillator are the fractional occupancy (by phosphates) of the $S^{431}$ and $T^{432}$ sites (Fig 2D). Assuming autonomous phosphorylation and dephosphorylation of the two sites, there is a simple mapping between the four phosphoform abundances and the occupancies of the two sites. As a result the experimental partial reaction kinetics for the phosphoforms can be fit with simple analytical functions in terms of these two state variables. Matrix simulations which include the complete KaiC heterogeneity in hexamer composition are used to study the two phase transitions with regard to the existence of oscillations and indicate a preferential role for the $S^{431}$ sites and in particular, serine-only monomers and monomer exchange, in initiating the dephosphorylation phase and synchronizing the return to phosphorylation.

Simulations of the clock suggest KaiB bound states initially form when serine-only monomers comprise at least 1/2 the hexamer (Fig 2C). These KaB-KaiC complexes sequester multiple KaiA dimers (3 to 4) and the resulting hexamer does not release the KaiA dimers until there are one or zero serine-only monomers remaining in the hexamer. Thus the proposed model suggests there is an asymmetry at the individual hexamer level in which an individual hexamer changes "state" with regard to KaiA sequestration, supporting the idea of at least two different conformational states of the hexamer (24). However in the model all individual hexamers and complexes remain phosphorylatable by free, non-sequestered KaiA so that individual hexamers do not change from a phosphorylating to dephosphorylating phase after KaiB association, but rather the rates across the population are determined by (global) non-sequestered KaiA . In the model KaiB can disocciate from the KaiABC complexes after initiating sequestration without affecting the dynamics and the conditions under which KaiB preferentially dissociates is in principle determinable from the kinetics of the various complexes. There is enhanced nonlinearity in the population phase transitions due to monomer exchange, and this also appears to explain the dominance of dephosphorylating phase samples under mixing of multiple samples (8). Therefore the simulations provide support for the basic mechanism proposed in Rust et al (16) consisting of serine monomer-associated KaiA sequestration where it has been demonstrated in this paper how the mechanism may operate at the level of protein-protein interactions. Finer tuning of the oscillatory period is possible in the model (and presumably in vivo and evolutionarily) via the KaiB association rate (binding probability in the model) and KaiA "release" rate. The general structure of the model also allows for a wide range of oscillatory behavior where mutations in KaiC could create lower thresholds for KaiB association and allow for periods at least as short as 10 hrs (Fig 2B). More generally the method of selective protein sequestration in complexes and selective release appears to be a viable method of obtaining temporal precision (synchrony) in biochemical processes.

The evidence for a cyclic scheme at the individual monomer level, (S,T) → (S,pT) → (pS,pT) → (pS,T), was inferred from the population phosphoform oscillations and the behaviour of phosphomimetic mutants of KaiC (15). We have shown that the population phosphoform dynamics for partial reactions and for the full system can be described by autonomous phosphorylation and dephosphorylation at $S^{431}$ and $T^{432}$ sites. The KaiC mutants do not share the characteristics of the WT KaiC protein and the matrix simulations have a heterogeneous mixture of monomers and require 3 serine-only monomers for KaiB binding while the mutants are mimicking hexamers with 6 serine-only monomers. For example, in nonlimiting KaiA the mutant data appear to predict that (pS,pT) → (pS,T) which is clearly not supported by the WT data as the system approaches ~80% phosphorylation (or larger for higher KaiA concentrations) asymptotically with a predominance (pS,pT) monomers. The analytical fits and model agreement suggest that a cycling scheme at the individual monomer level is not necessary, but the model cannot rule out cooperativity across adjacent sites.

## Methods
Matrix simulations were performed using an exectuble compiled from in-house code written in Fortran G77 (GNU, Free Software Foundation) as previously described (24). A 4th order Runge-Kutta solver was used in computing ODE solutions from in-house Fortran G77 exectuable code.

## Akcnowledgements
I would like to thank Carl Johnson, Tetsuya Mori and Ximing Qin at Vanderbilt U. for comments and suggestions on preliminary drafts of the MS. I would also like to thank C. Brettshneider and the Theoretical Biology Group at Humboldt U. for discussions related to the MS. This research was supported in part by a faculty development grant and Mitchell summer research grant from SHC .

## References

1. Nakajima M, Imai K, Ito H, Nishiwaki T, Murayama Y, et al. (2005) Reconstitution of circadian oscillation of cyanobacterial KaiC phosphorylation in vitro. Science 308: 414-415.

2. Tomita J, Nakajima M, Kondo T, Iwasaki H (2005) No transcription-translation feedback in circadian rhythm of KaiC phosphorylation. Science 307: 251–254

3. Nishiwaki T, Satomi Y, Nakajima M, Lee C, Kiyohara R, Kageyama H, Kitayama Y, Temamoto M, Yamaguchi A, Hijikata A, Go M, Iwasaki H, Takao T, Kondo T (2004) Role of KaiC phosphorylation in the circadian clock system of synechococcus elongatus PCC 7942. Proc Natl Acad Sci USA 101: 13927–1393.

4. Xu Y, Mori T, Pattanayek R, Pattanayek S, Egli M, Johnson CH (2004) Identification of key phosphorylation sites in the circadian clock protein KaiC by crystallographic and mutagenetic analyses. Proc Natl Acad Sci USA 101: 13933–13938

5. Kitayama Y, Nishiwaki T, Terauchi K, Kondo T (2008) Dual KaiC-based oscillations constitute the circadian system of cyanobacteria. Genes Dev 22: 1513-1521.

6. Zwicker D, Lubensky DK, ten Wolde PR (2010) Robust circadian clocks from coupled protein-modification and transcription–translation cycles. Proc Natl Acad Sci USA 107: 22540-22545.

7. Qin X, Byrne M, Xu Y, Mori T, Johnson CH (2010) Coupling of a core post-translational pacemaker to a slave transcription/translation feedback loop in a circadian system. PLoS Biol. 8, e1000394.

8. Ito H, Kageyama H, Mutsuda M, Nakajima M, Oyama T, et al. (2007) Autonomous synchronization of the circadian KaiC phosphorylation rhythm. Nat Struct Mol Biol 14: 1084-1088.

9. Mori T, Saveliev SV, Xu Y, Stafford WF, Cox MM, Inman RB, Johnson CH (2002) Circadian clock protein KaiC forms ATP-dependent hexameric rings and binds DNA. Proc Natl Acad Sci USA 99: 17203–17208

10. Nishiwaki T, Iwasaki H, Ishiura M, Kondo T (2000) Nucleotide binding and autophosphorylation of the clock protein KaiC as a circadian timing process of cyanobacteria. Proc Natl Acad Sci USA 97: 495–499.

11. Xu Y, Mori T, Johnson CH (2003) Cyanobacterial circadian clockwork: roles of KaiA, KaiB and the KaiBC promoter in regulating KaiC. EMBO J 22: 2117–2126.

12. Iwasaki H, Nishiwaki T, Kitayama Y, Nakajima M, Kondo T (2002) KaiA-stimulated KaiC phosphorylation in circadian timing loops in cyanobacteria. Proc Natl Acad Sci USA 99: 15788–15793.

13. Xu Y, Mori T, Johnson CH (2003) Cyanobacterial circadian clockwork: roles of KaiA, KaiB and the KaiBC promoter in regulating KaiC. EMBO J 22: 2117–2126.

14. Nishiwaki T, Satomi Y, Nakajima M, Lee C, Kiyohara R, Kageyama H, Kitayama Y, Temamoto M, Yamaguchi A, Hijikata A, Go M, Iwasaki H, Takao T, Kondo T (2004) Role of KaiC phosphorylation in the circadian clock system of synechococcus elongatus PCC 7942. Proc Natl Acad Sci USA 101: 13927–13932

15. Nishiwaki T, Satomi Y, Kitayama Y, Terauchi K, Kiyohara R, et al. (2007) A sequential program of dual phosphorylation of KaiC as a basis for circadian rhythm in cyanobacteria. EMBO J 26: 4029-4037.

16. Rust MJ, Markson JS, Lane WS, Fisher DS, O'Shea EK (2007) Ordered phosphorylation governs oscillation of a three-protein circadian clock. Science 318: 809-812.

17. Clodong S, Dühring U, Kronk L, Wilde A, Axmann I, Herzel H, Kollmann M (2007) Functioning and robustness of a bacterial circadian clock. Mol Syst Biol 3: 90

18. van Zon JS, Lubensky DK, Altena PRH, Rein ten Wolde P (2007) An allosteric model of circadian KaiC phosphorylation. Proc Natl Acad Sci USA 104: 7420–7425

19. Yoda M, Eguchi K, Terada TP, Sasai M (2007) Monomer-shuffling and allosteric transition in KaiC circadian oscillation. PLoS ONE 2: e408

20. Eguchi K, Yoda M, Terada TP, Sasai M (2008) Mechanism of robust circadian oscillation of KaiC phosphorylation in vitro. Biophys J 95:1773–1784

21. Nagai T, Tomoki P. Terada, and Masaki Sasai Synchronization of Circadian Oscillation of Phosphorylation Level of KaiC In Vitro. Biophys J. 2010 June 2; 98(11): 2469–2477. doi: 10.1016/j.bpj.2010.02.036

22. Qin X, Byrne M, Mori T, Zou P, Williams DR, McHaourab H, Johnson CH (2010) Intermolecular associations determine the dynamics of the circadian KaiABC oscillator. Proc. Natl. Acad. Sci. USA 107, 14805-14810.

23. Brettschneider C, Rose RJ, Hertel S, Axmann I, Albert, Heck JR, and Kollmann M (2010) A sequestration feedback determines dynamics and temperature entrainment of the KaiABC circadian clock. Mol Syst Biol. 6: 389.

24. Mori T, Williams DR, Byrne MO, Qin X, Egli M, et al. (2007) Elucidating the ticking of an in vitro circadian clockwork. PLoS Biol 5: e93.

25. Kageyama H, Nishiwaki T, Nakajima M, Iwasaki H, Oyama T, et al. (2006) Cyanobacterial circadian pacemaker: Kai protein complex dynamics in the KaiC phosphorylation cycle in vitro. Mol Cell 23: 161-171.

26. Nakajima M, Ito H, Kondo T (2010) In vitro regulation of circadian phosphorylation rhythm of cyanobacterial clock protein KaiC by KaiA and KaiB. FEBS Lett 584(5): 898–902.

**Figure Legends**

**Figure 1.** A two-state model fits the experimental phosphorylation and dephosphorylation phase phosphoform kinetics (15,16).
**Panel A.** Phosphoform analytical curves based on the equations in the text for threonine-only monomers (green), doubly phosphorylated monomers (blue), and serine-only monomers (red). For the phosphorylation phase no KaiB is included (KaiA+KaiC only, $[KaiA_0]/[KaiC_0] =1$). At t = 20 hours the fraction of free KaiA was set to zero ($f_A =0$ for t > 20 hrs) and the system dephosphorylates. Both the analytical solutions from the text and a stochastic matrix simulation with N = 2000 individual hexamers are shown. The black curve is the simple sum of the monomer curves for comparison with Ref 16.
**Panel B.** The same curves in Panel A are equivalently described in terms of two states: the fractional occupancy of the $S^{431}$ sites and $T^{432}$ sites in the population.
**Panel C.** The distribution of phosphates across the KaiC hexamer population at the end of the phosphorylation phase (~20 hrs) from matrix simulations; the results from two different simulations are shown.

**Figure 2.** Simulations constrained by the partial reaction rates indicate the relevance of serine-only monomers for oscillations.
**Panel A.** Net percent phosphorylation of sites in the population $(f_S+f_T)/2$ versus time (hrs). KaiB binding and KaiA release independent of KaiC hexamer phosphoform composition does not produce sustained circadian oscillations. A varying threshold from 10 phosphates (highest curve maximum) to 5 phosphates (lowest curve maximum) for KaiB binding to a hexamer is shown. KaiA and KaiB are released when a hexamer completely dephosphorylates.
**Panel B.** Circadian oscillations are consistent with binding of KaiB to KaiC hexamers with a minimum threshold of three serine-only monomers. The net percent phosphorylation is shown for a varying threshold from 1-4 serine-only monomers. Thresholds of five and six serine-only monomers produce similar non-oscillatory curves as a threshold of four. KaiA and KaiB are released when a hexamer has zero serine-only monomers.
**Panel C.** The simulated phosphoform kinetics using the best candidate model from Panel B which is in good agreement with the experimental data from Refs. 15 and 16 .
**Panel D.** A simple 2-state deterministic ODE model with KaiA sequestration can describe circadian phosphoform oscillations (see text for model). The determinsitic model mimics the matrix model but does not include explicit KaiB binding and KaiA release or monomer exchange.

**Figure 3.** The matrix model is consistent with experimental constraints for the standard in vitro concentrations.
**Panel A.** Simulated abundances of complexes (as a fraction if initial KaiC) versus time. The serine monomer kinetics is shown for phase comparison.
**Panels B and C.** Simulated mixing of two oscillatory KaiC population samples with varying phase. In each case the dephosphosphorylating phase sample is dominant as described experimentally (8).

**Figure 4.** The matrix model can be used to interpret the kinetics for varying initial concentrations of proteins.
**Panel A.** Simulated effect of varying the initial KaiA concentration keeping KaiB and KaiC fixed at the standard concentrations.
**Panel B.** Simulated effect of varying the initial KaiB concentration keeping KaiA and KaiC fixed at the standard concentrations.
**Panel C.** Simulated effect of varying all three proteins simultaneously; see text for details.

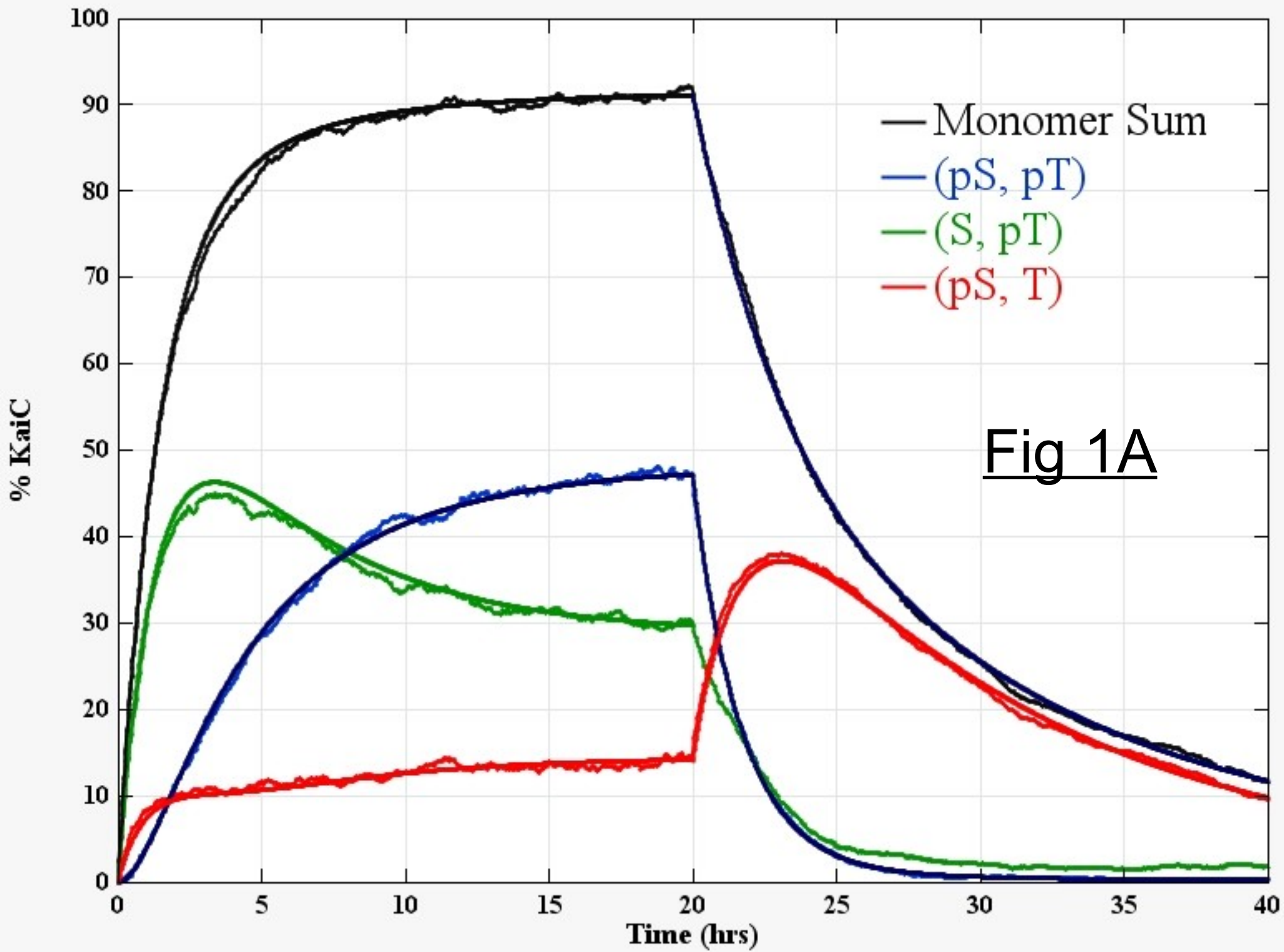

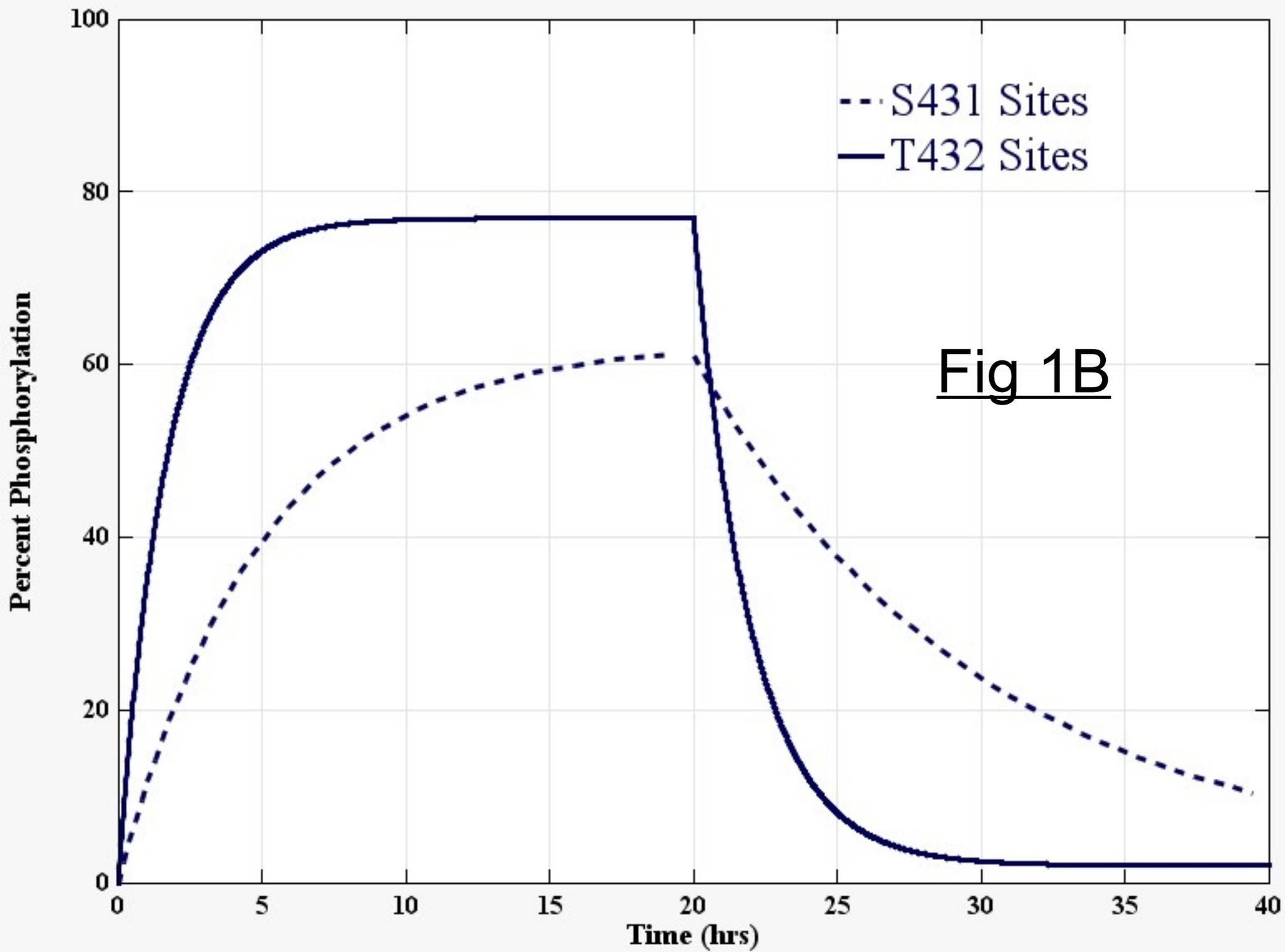

Fig 1B

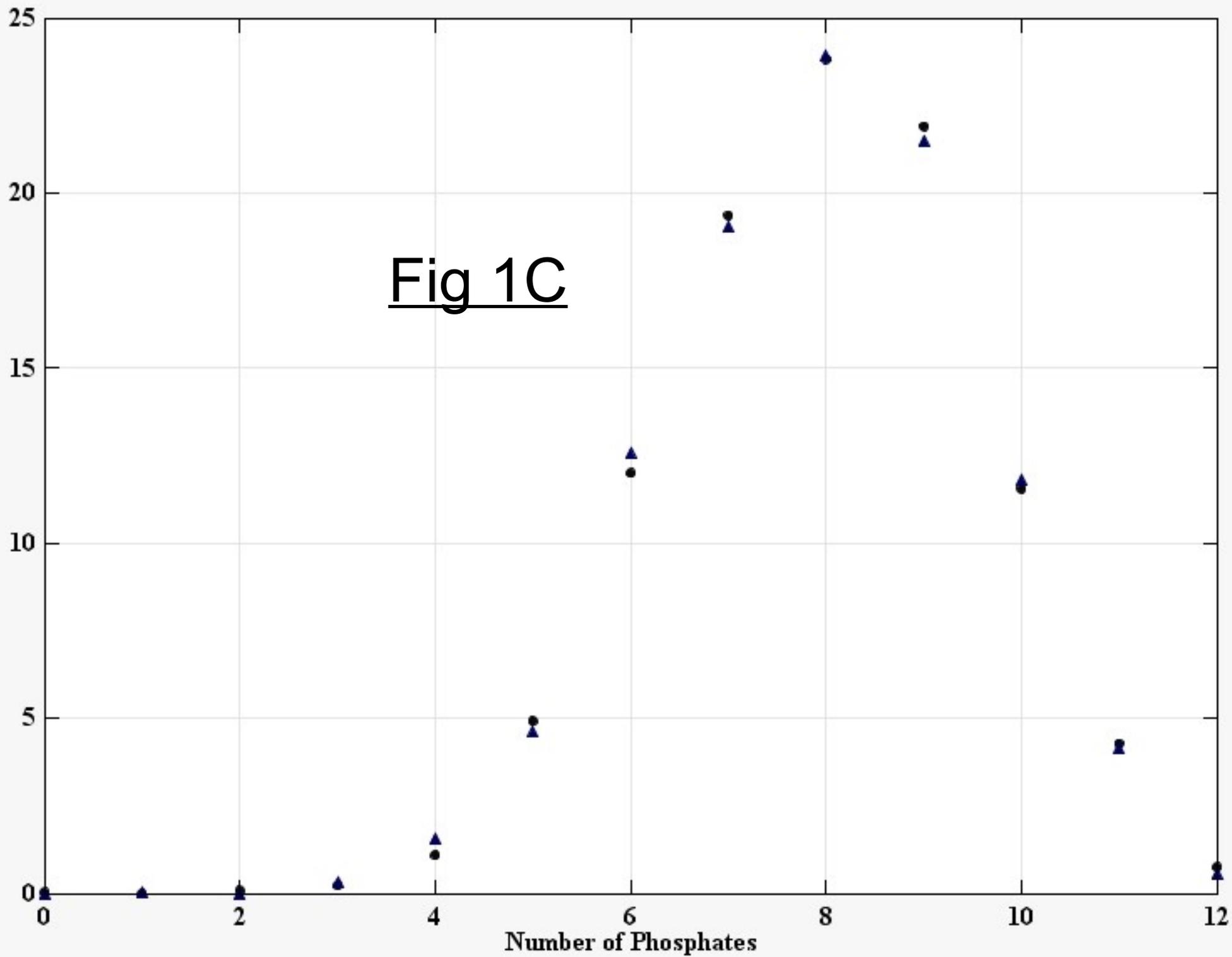

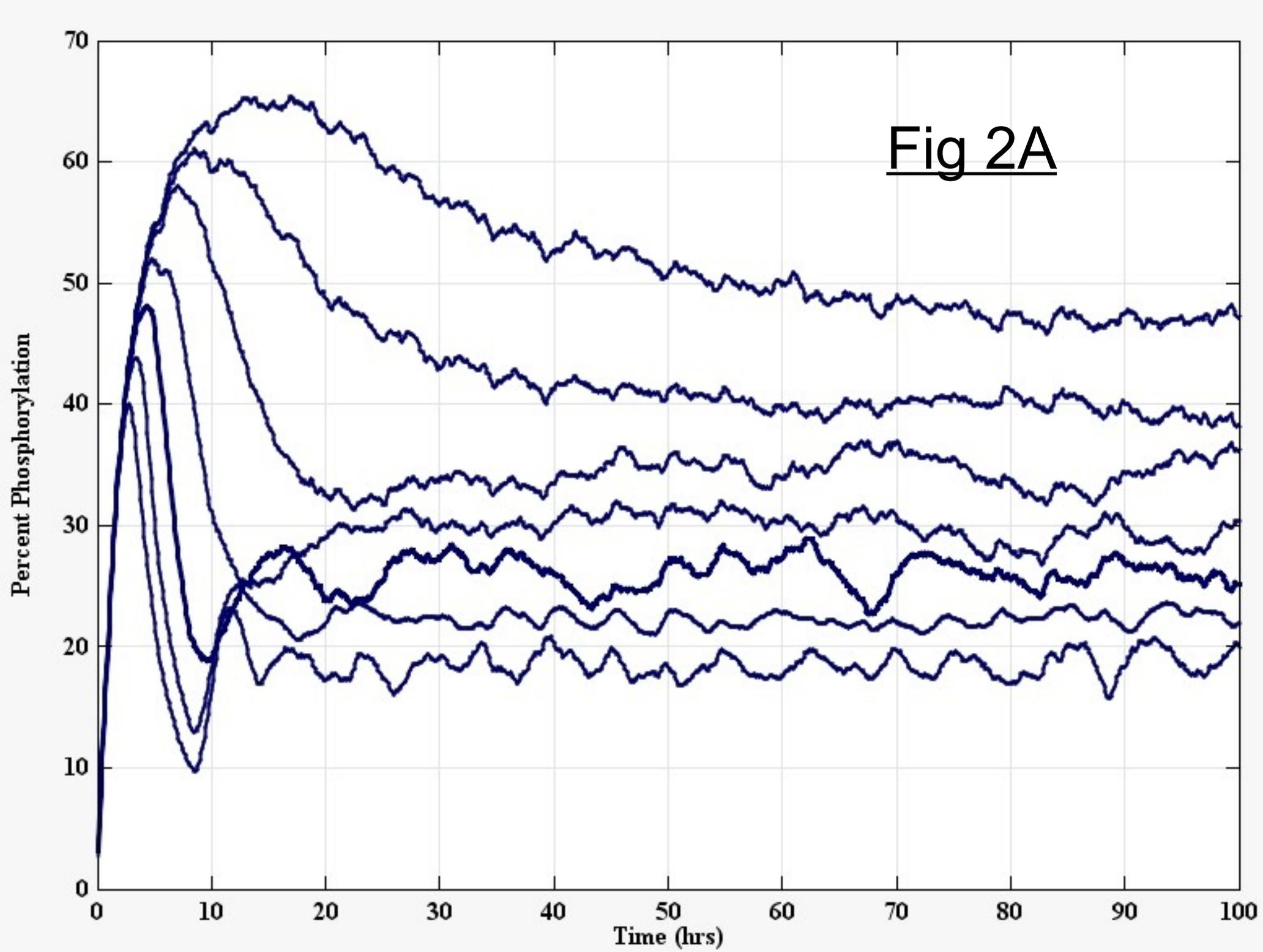
Fig 2A

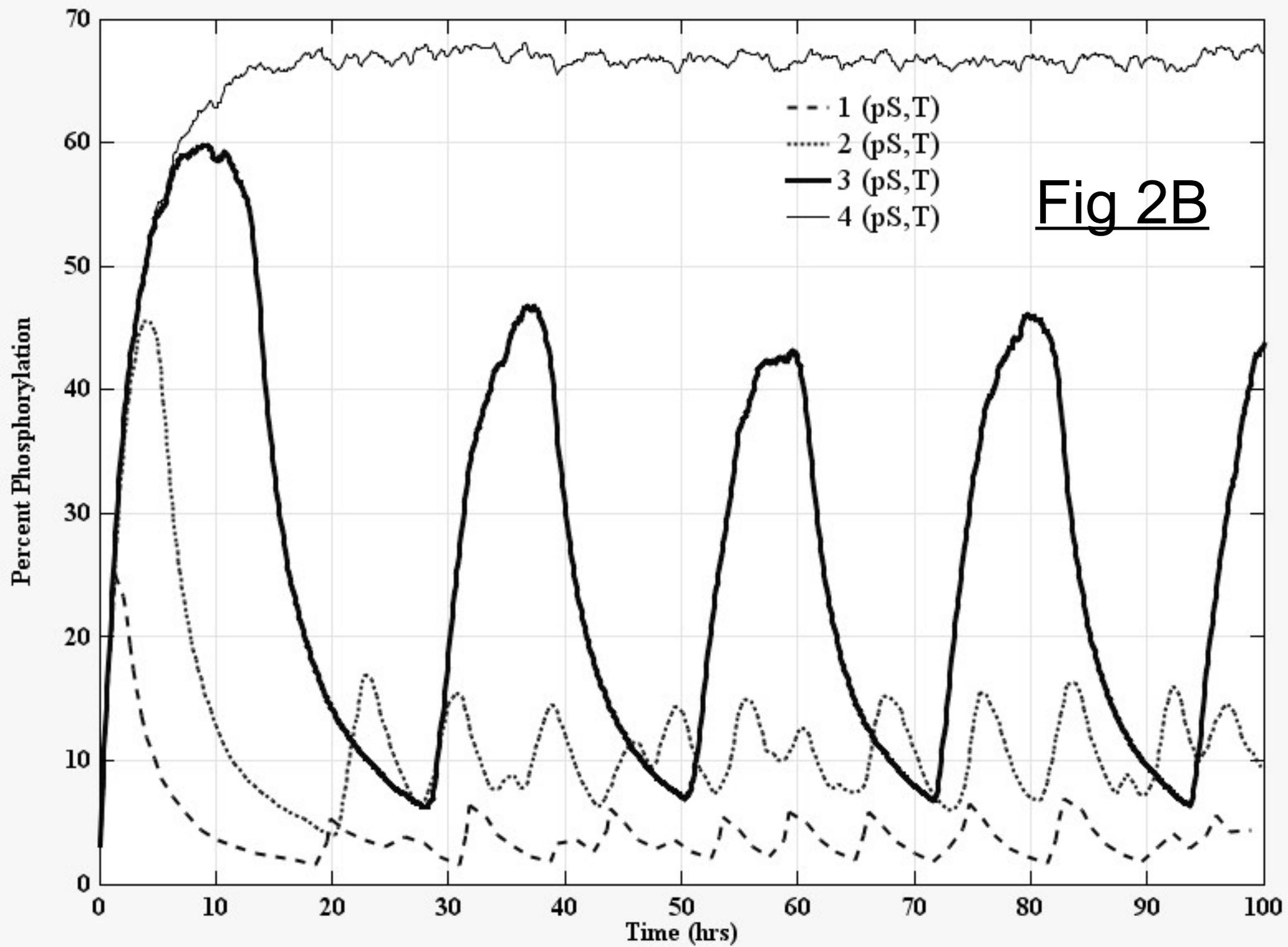

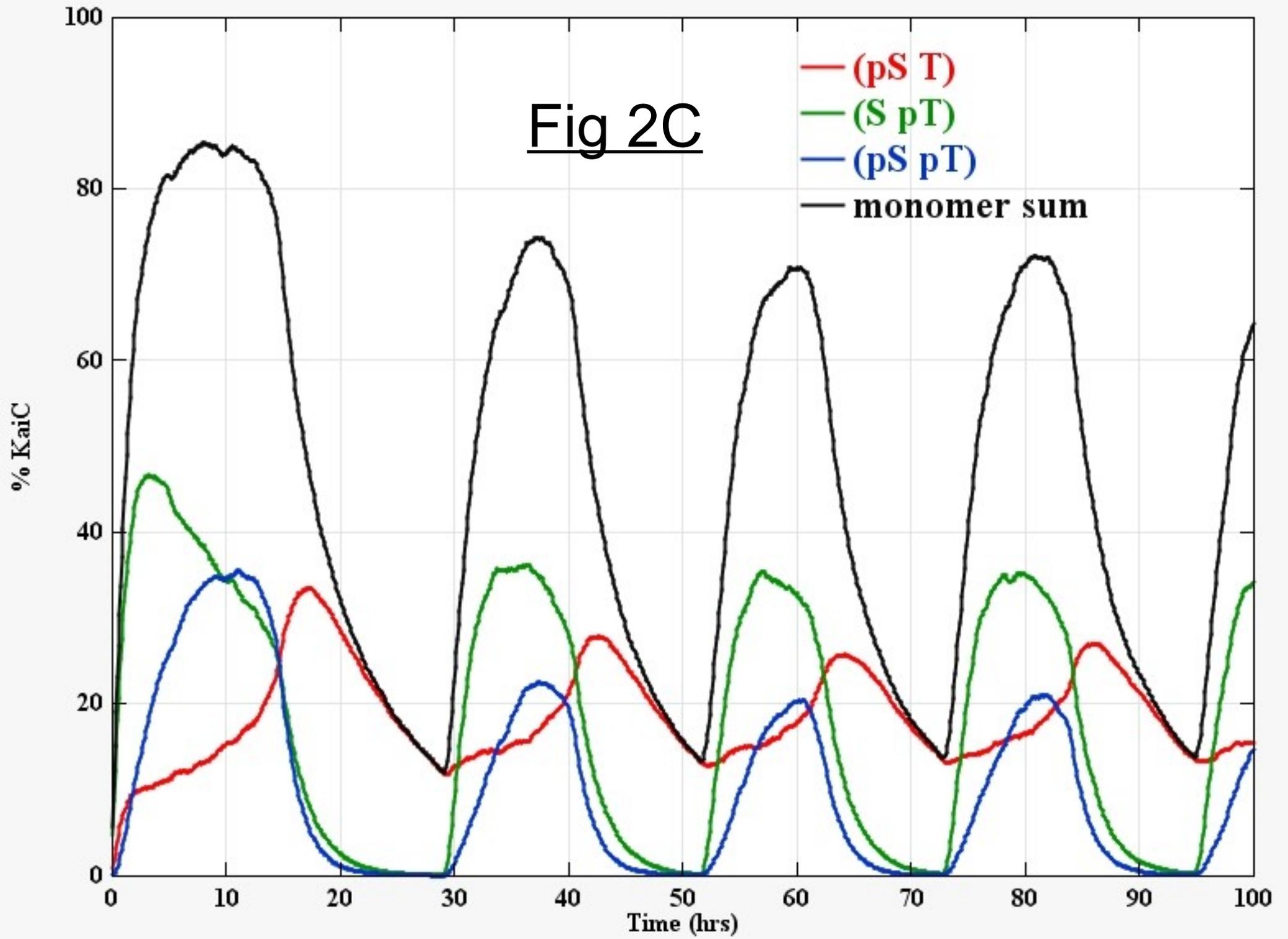

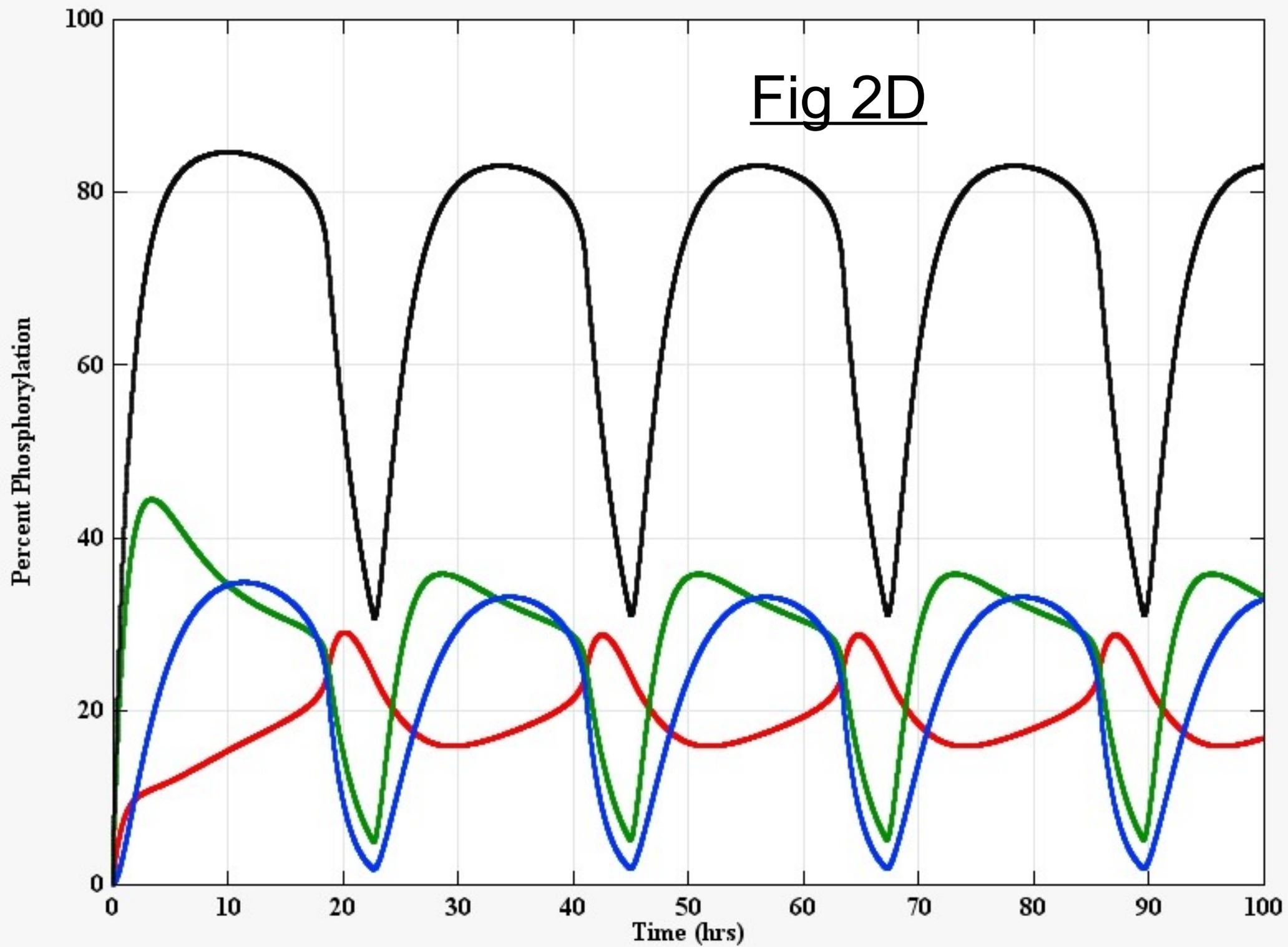

Fig 2D

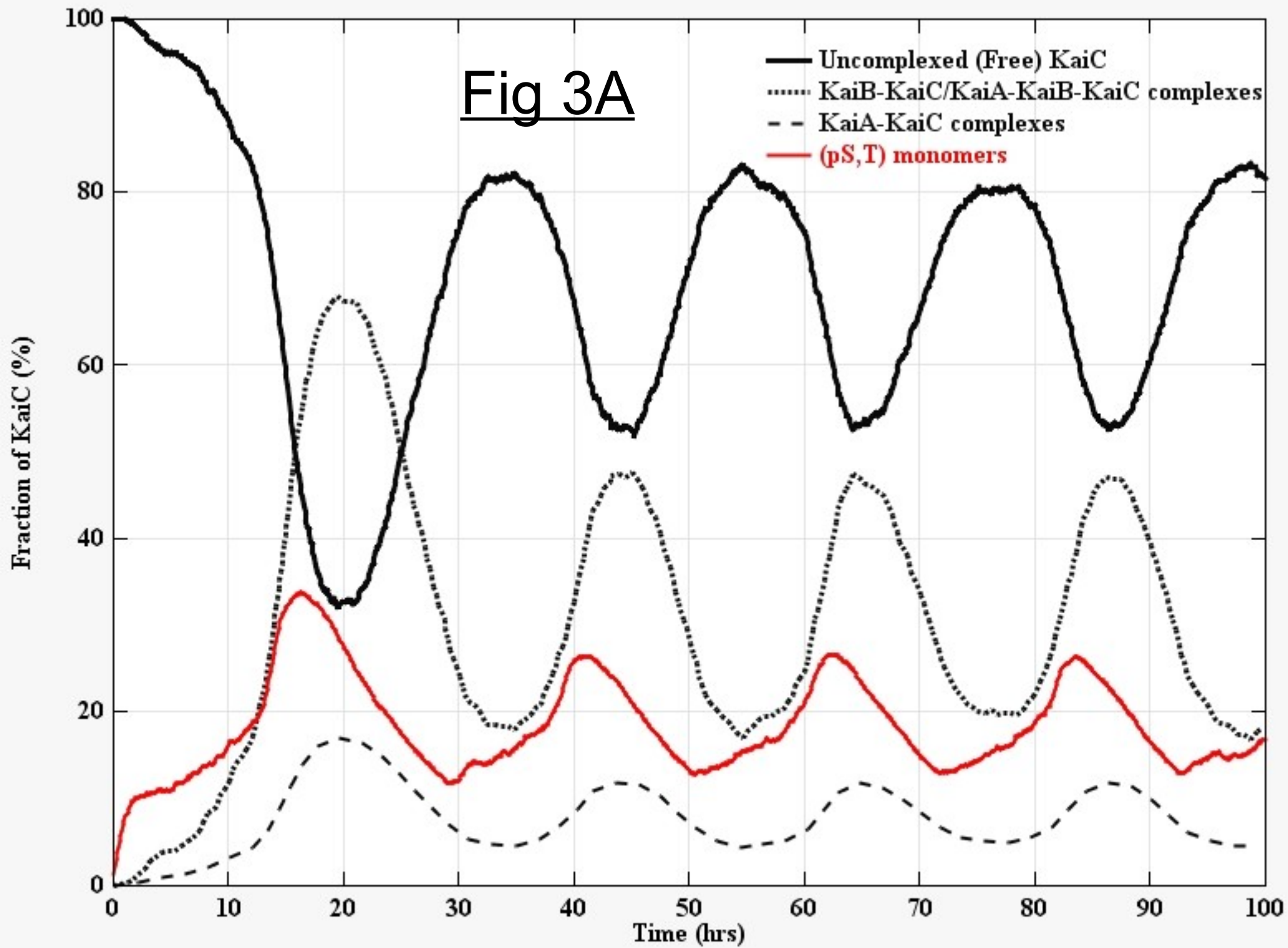

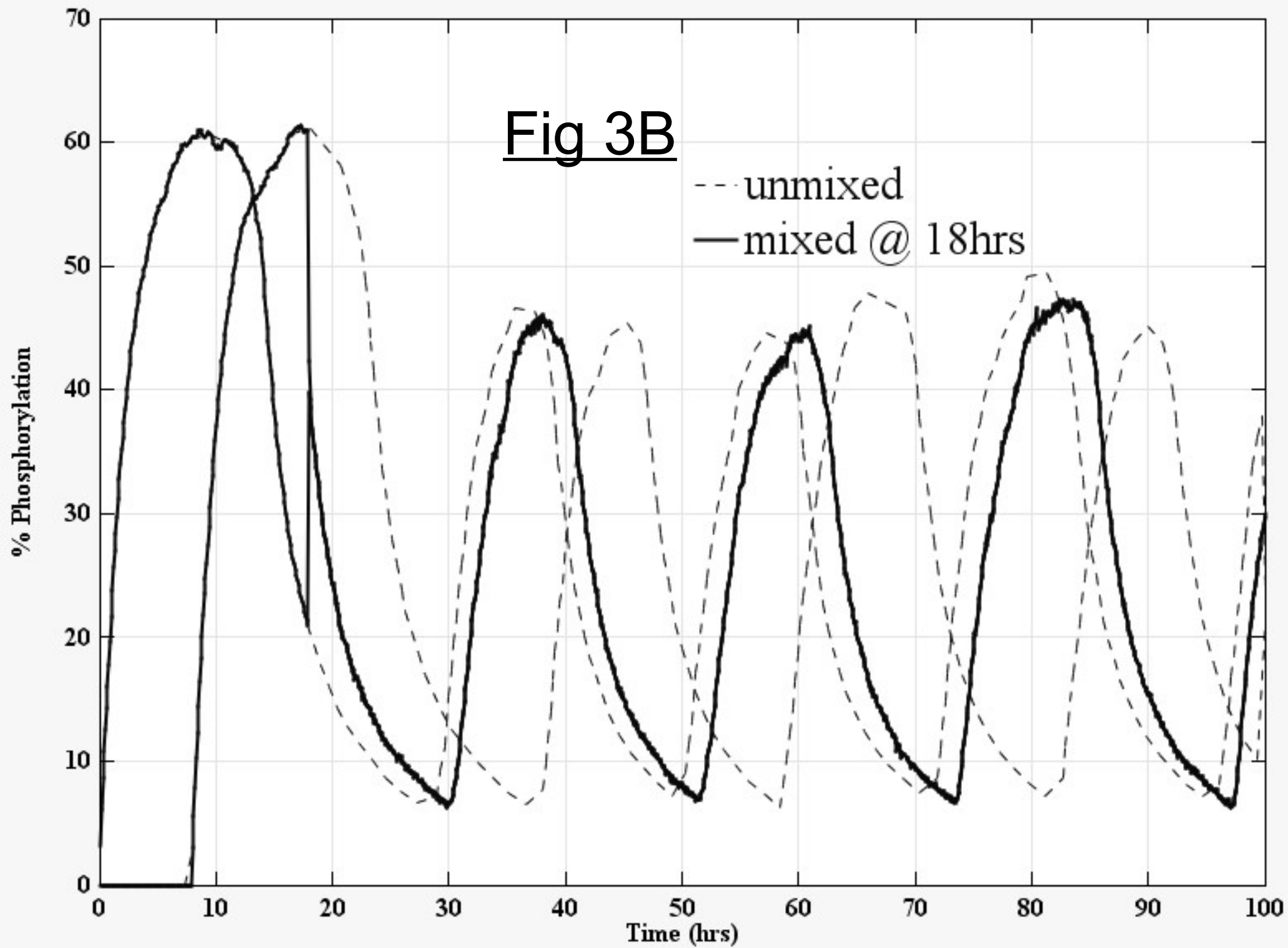

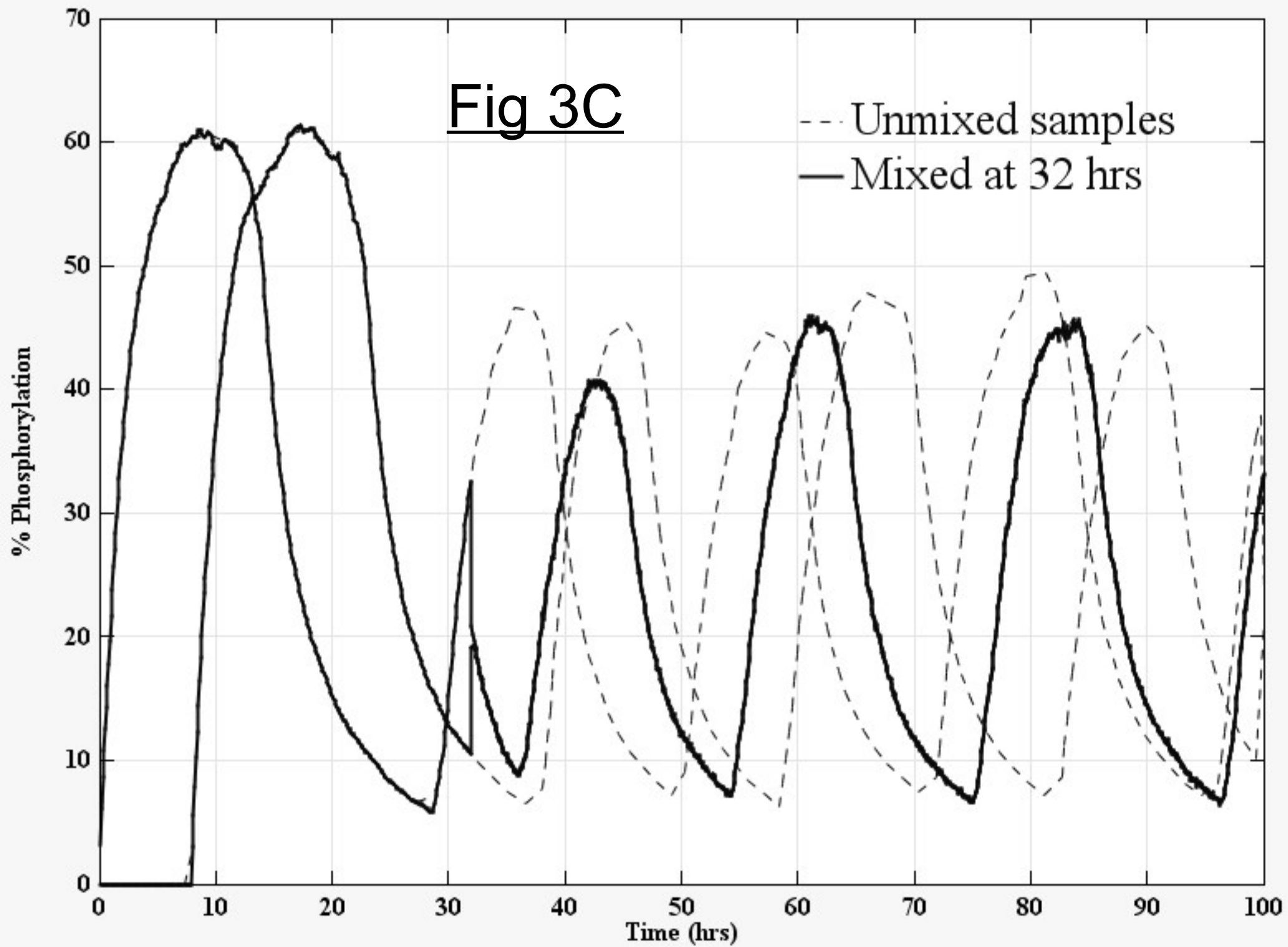

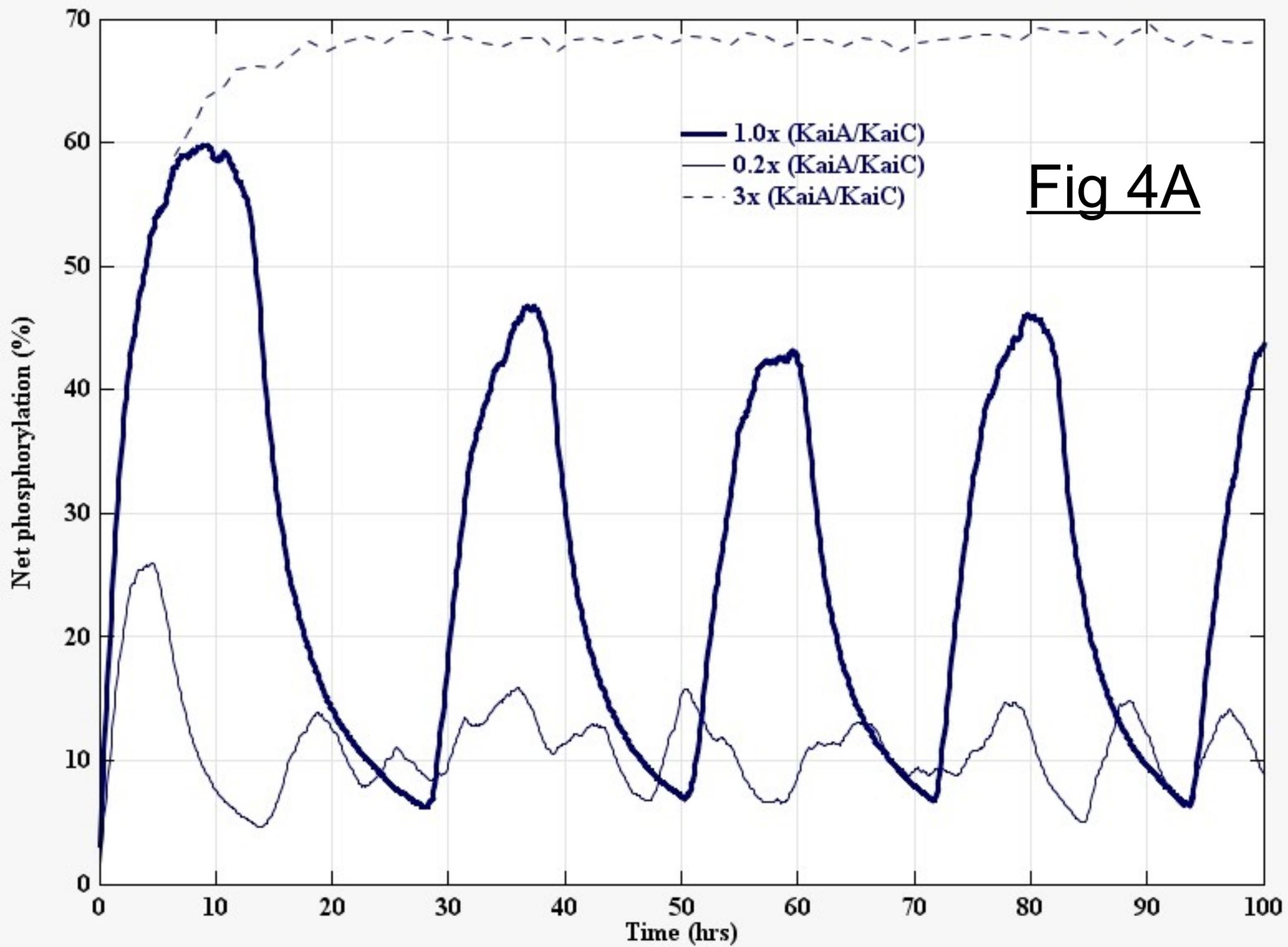

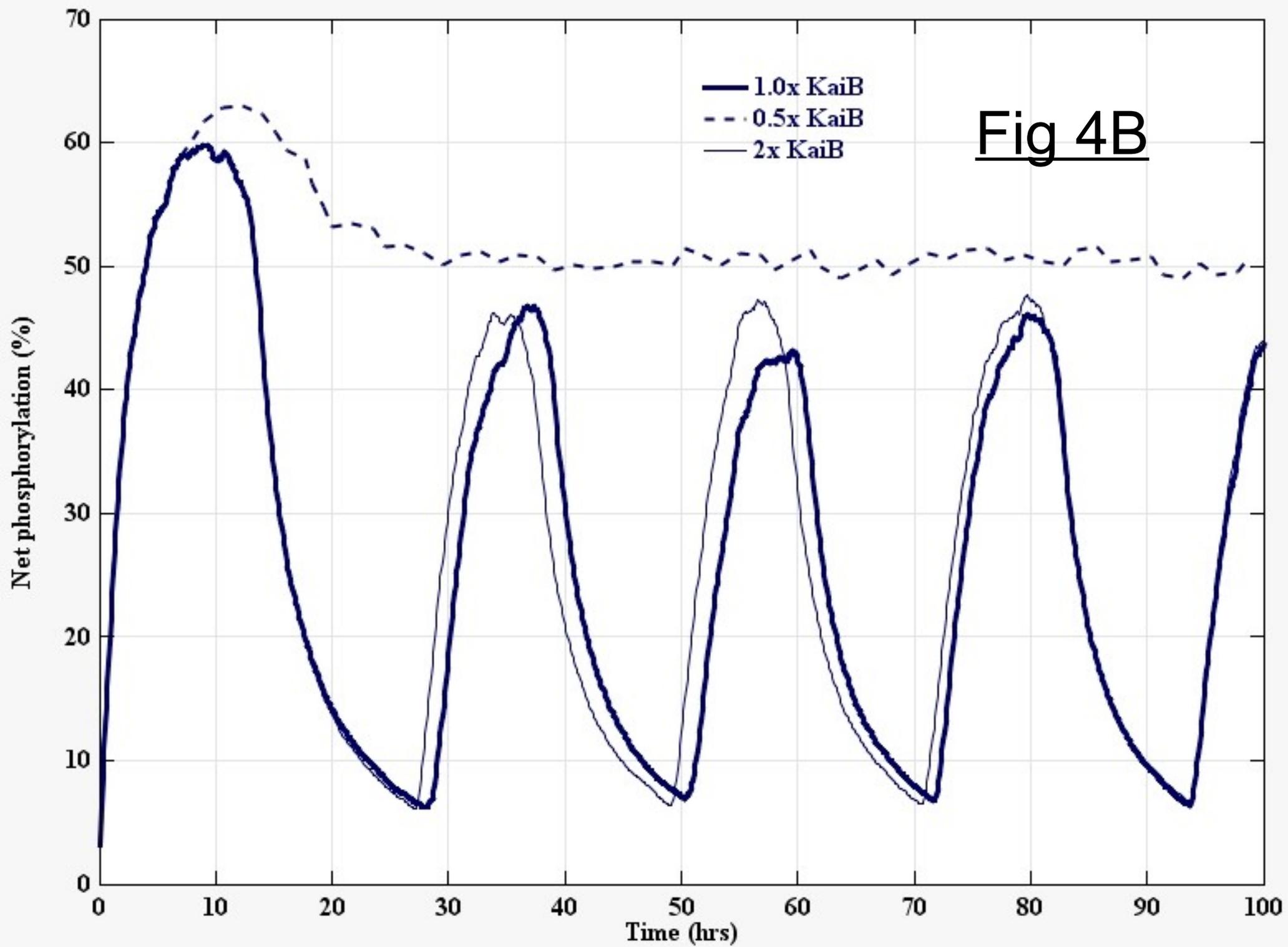

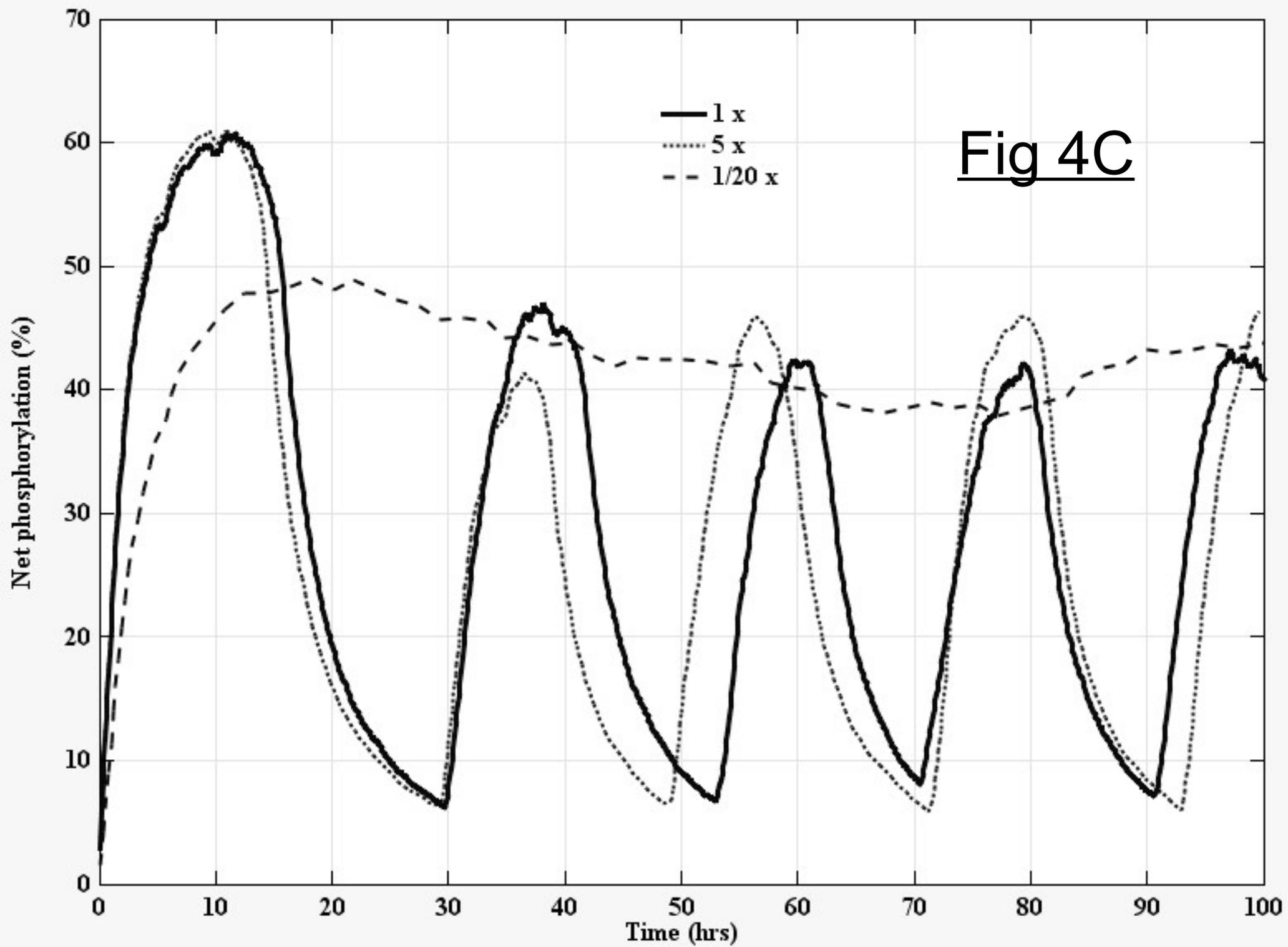
Fig 4C

**SI Text**

**I. Derivation of analytical phosphorm kinetics and fits to experimental data.**
Let $N_T(t)$ represent the total number of sites phosphorylated at T432 in the population and $N_S(t)$ the total number of sites phosphorylated at S431. Similarly let $N_{UT}$ be the total number of unphosphorylated T432 sites and $N_{US}$ be the total number of unphosphorylated S431 sites. If the sites are independent with regard to phosphorylation and dephosphorylation reactions,

(1) $\qquad dN_T/dt = k_{+T} N_{UT} - k_{-T} N_T$
(2) $\qquad dN_S/dt = k_{+S} N_{US} - k_{-S} N_S$

The rate $k_+$ refers to phosphorylation and $k_-$ to the rate of dephosphorylation, using a subscript T for the T432 sites and subscript S for the S431 sites. The total number of S and T sites is constant in time in the in vitro reactions: $N_{UT} + N_T = \text{constant} \equiv N_{max,T}$; $N_{US} + N_S = \text{constant} \equiv N_{max,S}$. Replacing $N_{UT}$ in equation (1) it follows that

(3) $\qquad df_T/dt = k_{+T}(1 - f_T) - k_{-T} f_T$

where $f_T \equiv N_T/N_{max,T}$ are the fraction of T sites phosphorylated. Similarly defining $f_S \equiv N_S/N_{max,S}$:

(4) $\qquad df_S/dt = k_{+S}(1 - f_S) - k_{-S} f_S$

If the effective rates are constant in time the solution to (3) starting with a completely unphosphorylated population is

(5) $\qquad f_T(t) = [1/(1 + k^A_{-T}/k^A_{+T})]\{1 - \exp[-(k^A_{+T} + k^A_{-T})t]\}$ $\qquad$ (KaiA-KaiC)

These rates assume some fixed KaiA concentration in a population model (and no KaiB) and make the approximation that this rate does not change as the hexamers phosphorylation level varies (e.g., due to possible differential affinity of KaiC for KaiA as a function of KaiC phosphorylation). The T432 phosphorylation occupancy should asymptotically rise to a maximum value of $[1/(1 + k_{-T}/k_{+T})]$ which allows a determination of the ratio of rates (for a fixed KaiA concentration). The time-constant for the rise is

(6) $\qquad \tau_T = 1/(k^A_{+T} + k^A_{-T})$ $\qquad$ (KaiA-KaiC)

which, together with the asymptotic value allows for the determination of the phosphorylation and dephosphorylation rates. The same analysis holds for the S sites (for the same initial conditions):

(7) $\qquad f_S(t) = [1/(1 + k^A_{-S}/k^A_{+S})]\{1 - \exp[-(k^A_{+S} + k^A_{-S})t]\}$ $\qquad$ (KaiA-KaiC)

with a time constant

(8) $\qquad \tau_S = 1/(k^A_{+S} + k^A_{-S})$ $\qquad$ (KaiA-KaiC)

Since, experimentally, the phosphorylation rate of S431 is less than T432 the S431 time constant is rate-determining for oscillations. Similarly for the dephosphorylation phase (KaiB-KaiC or KaiC alone) we assume constant rates and starting with an initial hyper-phosphorylated fraction $f_{0,T}$ we find

the dephosphorylation kinetics from Eqn (3):

(9) $\quad f_T(t) = [1/(1 + k_{-T}/k_{+T})] \{ 1 - [1 - f_{0,T}(1 + k_{-T}/k_{+T})] \exp[-(k_{+T} + k_{-T})t] \}$ (KaiC alone, KaiB-KaiC)

The asymptotic value is in terms of the ratio of the rates, $f_T(t \to \infty) = [1/(1 + k_{-T}/k_{+T})]$. There is an analogous solution for the fraction of S431 sites phosphorylated.

The relations above describe the kinetics of the degree of phosphorylation of individual sites (S and T) in the population, but not the monomer kinetics. If placing these phosphates randomly into monomers we will find the expected fraction of S-only monomers (pS,T) by considering the product of the probability that T is *not* phosphorylated (with probability $(1 - f_T)$) and the probability that S is phosphorylated (with probability $f_S$). The fraction of S-only monomers is therefore $f_{(pS,T)} = f_S*(1 - f_T)$. Similarly the fraction of ST monomers is $f_{(pS,pT)} = f_S*f_T$ and T-only monomers is $f_{(S,pT)} = f_T*(1 - f_S)$.

For the KaiA-KaiC partial reaction the site kinetics are given by the simple exponential solutions, equations (5) and (7). To make the expressions more tractable let us define $\beta = k_-/k_+$ and $\alpha = k_+ + k_-$ using subscripts and superscripts to indicate the site and whether the rate refers to KaiA. Then we find for the KaiA-KaiC "phosphorylation" phase the predicted analytical curves for the monomer abundances:

(10) $\quad f_{(S,pT)} = f_T*(1 - f_S) = [1/(1 + \beta^A_T)]*[1/(1 + \beta^A_S)] (1 - \exp[-\alpha^A_T t])*(\beta^A_S + \exp[-\alpha^A_S t])$
(11) $\quad f_{(pS,T)} = f_S*(1 - f_T) = [1/(1 + \beta^A_T)]*[1/(1 + \beta^A_S)] (1 - \exp[-\alpha^A_S t])*(\beta^A_T + \exp[-\alpha^A_T t])$
(12) $\quad f_{(pS,pT)} = f_S*f_T = [1/(1 + \beta^A_T)]*[1/(1 + \beta^A_S)] (1 - \exp[-\alpha^A_S t])*(1 - \exp[-\alpha^A_T t])$

The asymptotic value of (pS,pT) determines $[1/(1 + \beta^A_T)]*[1/(1 + \beta^A_S)] = f_{(pS,pT)}(t \to \infty)$. The asymptotic value of (pS,T) from experiment determines $\beta^A_T$ since $f_{(pS,T)}(t \to \infty) = \beta^A_T * f_{(pS,pT)}(t \to \infty)$. The asymptotic value of (S,pT) determines $\beta^A_S$ since $f_{(S,pT)}(t \to \infty) = \beta^A_S * f_{(pS,pT)}(t \to \infty)$. A fit to the time dependence fixes the values of $\alpha$.

Asymptotic values (end of phosphorylation phase):
$f_{(pS,pT)} \to 0.4$ and $f_{(pS,T)} \to 0.15$ and $f_{(S,pT)} \to 0.25$ (approximately)
implies $\beta^A_T \sim 0.15/0.4 = 0.35$ ; $\beta^A_S \sim 0.25/0.4 = 0.62$

Implies approximate fitted rates (KaiA+KaiC, standard concentrations, hr$^{-1}$):
$k^A_{+S} \sim 0.12$, $\quad\quad\quad k^A_{-S} \sim 0.076$
$k^A_{+T} \sim 0.444$ $\quad\quad\quad k^A_{-T} \sim 0.155$

We can also fit analytical expressions in the two-state model to the dephosphorylation phase, where in the partial reaction data KaiA was removed (immunoprecipitation) after each state reached their approximate pseudo-asymptotic values. To simplify further let $\gamma = 1 + k_-/k_+ = 1 + \beta$. For each site during dephosphorylation its time-dependence is given by

(13) $\quad f(t) = (1/\gamma) \{ 1 - [1 - \gamma f_0] \exp[-\alpha t] \}$

where $f_0$ is the initial fraction of that site prior to starting dephosphorylation. Applying this to the monomers:

(14) $\quad f_{(S,pT)} = f_T*(1 - f_S) = (1/\gamma_T) (\{ 1 - [1 - \gamma_T f_{0T}] \exp[-\alpha_T t] \} * (1 - (1/\gamma_S) * \{ 1 - [1 - \gamma_S f_{0S}] \exp[-\alpha_S t] \})$

(15) $f_{(pS,T)} = f_S*(1-f_T) = (1/\gamma_S) ( \{ 1 - [1 - \gamma_S\ f_{0S}]\ exp\ [-\alpha_S\ t]\ \}*( 1 - (1/\gamma_T) *\{ 1 - [1 - \gamma_T\ f_{0T}]\ exp\ [-\alpha_T\ t]\ \}\ )$

(16) $f_{(pS,pT)} = f_S*f_T = (1/\gamma_S) (1/\gamma_T) \{ 1 - [1 - \gamma_S\ f_{0S}]\ exp\ [-\alpha_S\ t]\ \}*\{ 1 - [1 - \gamma_T\ f_{0T}]\ exp\ [-\alpha_T\ t]\ \}$

The parameters $\beta_S = 10$, $\beta_T = 30$, $\alpha_S = 0.1$ hr$^{-1}$, and $\alpha_T = 0.5$ hr$^{-1}$ produce a good fit to the dephosphorylation phase kinetics. Again it appears that the two-state model is likely sufficient to describe these partial reactions. The apparent "transition" to the (pS,T) state is really a consequence of fast T432 site dephosphorylation versus slower S431-site dephosphorylation.

The kinetics of dephosphorylation constrain $\alpha$; $k_- \gg k_+$ $k_{-T} \sim 0.5$ and $k_{-S} \sim 0.1$ dephosphorylation rates (hr$^{-1}$) for the dephosphorylation phase. However there is very little constraint on the ratio of rates ($\beta$) in either case except they are $k_-/k_+ \gg 2$; this ratio could be very high. In simulations we have set $k_{+S}$ and $k_{+T}$ to zero (in the dephosphorylation phase) for simplicity.

**II. Deterministic "2-state" model (Fig 2D)**

(17)
$$df_S/dt = k_{+S}(1-f_S) - k_{-S} f_S$$
$$df_T/dt = k_{+T}(1-f_T) - k_{-T} f_T$$

$$k_i = k_i^0 + k_i^A f_A/(f_A + K')$$

In the parameterization of Eqn (17):

rates, T432: $k_{+,T}^0 = 0; k_{-T}^0 \approx 0.5; k_{+T}^A \approx 0.5; k_{-T}^A \approx -0.4$
rates, S431: $k_{+,S}^0 = 0; k_{-S}^0 \approx 0.15; k_{+S}^A \approx 0.14; k_{-S}^A \approx -0.08$

**Implies:**
$$k_{+,S} \approx 0.1 f_A/(f_A + K')\ ;\ k_{-,S} \approx 0.1 - 0.08 f_A/(f_A + K')$$

$$k_{+,T} \approx 0.5 f_A/(f_A + K')\ ;\ k_{-,T} \approx 0.5 - 0.36 f_A/(f_A + K')$$

K' = 0.13     (K' = 0.3 in matrix simulations)
$f_A(t) = f_{Ai} - N_s*f_{(pS,T)} = max\{0, f_{Ai} - N_s*[f_S(1-f_T)]\}$
$N_s \sim 4.1$     ($N_s = 3.5$ in matrix simulations)

The eight rates from the analytical fits are used in the parameterization of Eqn (17) as follows. For the KaiA+KaiC partial reaction we use $f_A = [A]/[C] = 1$ (dimer/hexamer):

$$k_i^A = (1 + K')k_{i,fit}^A - k_{i,fit}^0$$
$$k_i^0 = k_{fit}^0$$

For concentration changes in KaiB (Fig 4B) the following association rate (per molecule per hr) to hexamer KaiC was assumed: $k_{+B} = 0.75 f_B/(f_B + 0.1)$ where $f_B \equiv [B]/[B_{standard}]$.